# 64-qubit quantum circuit simulation

Zhao-Yun Chen[1,2][†], Qi Zhou[1,2][†], Cheng Xue[1,2], Xia Yang[2], Guang-Can Guo[1], Guo-Ping Guo[1,*]

[†] These authors contributed equally to this work.

1. Key Laboratory of Quantum Information, University of Science and Technology of China, CAS, Hefei 230026, China.
2. Origin Quantum Computing Company Limited, Hefei 230026, China.

*Corresponding author (email: Guo-Ping Guo, gpguo@ustc.edu.cn)

**Abstract**
Classical simulations of quantum circuits are limited in both space and time when the qubit count is above 50, the realm where quantum supremacy reigns. However, recently, for the low depth circuit with more than 50 qubits, there are several methods of simulation proposed by teams at Google and IBM. Here, we present a scheme of simulation which can extract a large amount of measurement outcomes within a short time, achieving a 64-qubit simulation of a universal random circuit of depth 22 using a 128-node cluster, and 56- and 42-qubit circuits on a single PC. We also estimate that a 72-qubit circuit of depth 23 can be simulated in about 16 h on a supercomputer identical to that used by the IBM team. Moreover, the simulation processes are exceedingly separable, hence parallelizable, involving just a few inter-process communications. Our work enables simulating more qubits with less hardware burden and provides a new perspective for classical simulations.

# 1. Introduction

The last few years have seen a series of significant advances in quantum computing, in particular regarding superconducting quantum chips with reports of devices of 20 and 50 qubits with good fidelity [1-2]. In the meantime, great progress has also been made with semiconductor quantum chips [3-5]. "Quantum supremacy" claims that the limit of classical computers would be transcended if a device of 50 qubits were made [6]. Direct simulations of 50 qubits take about 16-PB of RAM to store the full vectors. Google and IBM teams have proposed some efficient methods for simulating the low-depth circuit with more than 49 qubits (e.g., deferral of entanglement gates [7] and Feynman path method [8]). Here, we present a scheme to optimize the classical simulation of quantum circuits with low depth and large sampling number, with which we have performed a 64-qubit simulation with sampling number $2^{28}$ and depth 22. In particular, by transforming several control-Z (CZ) gates to measurement and single-qubit gates, the circuit is mapped onto an additional $2^n$ sub-circuits. These sub-circuits are formed by two blocks without any qubit entanglement between them, thereby converting an $N$ qubit simulation problem into a group of $N/2$. Our method is similar to a small balanced cut in a two-dimensional grid [9], while the method developed by Aaronson is more general but more complicated as a compromise. For decomposing one CZ gate, their method splits the original circuit into eight sub-circuits, while in our case it splits into four. The results of all the sub-circuits are then added together to reconstruct the final state. In practice, we simulated the universal random circuit, which is used to characterize the quantum supremacy in the region of quantum chaos [10-17].

# 2. Methods

*2.1 partition scheme*

A CZ gate can be transformed into two groups of measurement and single-qubit gates, specifically

$$CZ = P_0 \times I + P_1 \times Z, \quad (1)$$

where $P_0 = \begin{pmatrix} 1 & 0 \\ 0 & 0 \end{pmatrix}$, $P_1 = \begin{pmatrix} 0 & 0 \\ 0 & 1 \end{pmatrix}$. $I$ denotes the unit matrix, and $Z$ the Pauli-$Z$ matrix.

The transformation dismisses the entanglement gate between the two qubits and makes a copy to the circuit. We illustrate an 8-qubit circuit of depth 8 as an example (Fig. 1). By transforming CZ gates in the seventh and eighth layer (in the dashed boxes), the original circuit is converted to four copies. The final state is the addition of the four final states of each copy. In the four copies, qubits 0 to 3 are no longer entangled with qubits 4 to 7. Hence we can simulate them separately. At the beginning, 8 qubits can represent $2^8$ states, and after the conversion, there are 8 circuits each with 4 qubits representing $8 \times 2^4$ states (see the grey dashed boxes in the bottom of Fig. 1); the space is reduced. Initially, there are 27 gates of 8 qubits, and after the conversion, there are 112 gates of 4 qubits — the time is therefore reduced.

In practice, we divided each of the 4 qubit into two half-circuits — an upper and a lower part as shown in the bottom of Fig. 1. In the first six layers, there is no CZ gate entangling the two half-circuits. In the seventh and eighth layers, the CZ gates entangling the two half-circuits should be transformed as mentioned above. The same holds for the transformation executed on the 15$^{th}$ and 16$^{th}$ layers. This generates $2^c$ copies to be simulated, where $c$ is the number of transformed CZ gates.

## 2.2 Methods to estimate the different qubit counts and depths

We provide in Table 1 a set of time estimates for various qubit topologies and circuit depths. The time estimation are obtained from

$$\text{Time} = \sum_{i=1}^{d(\text{depth})} n_i(\text{gates}) \times m(\text{circuits}) \times t(\text{time / gate}) / s(\text{nodes}), \qquad (2)$$

where $n_i$ is the number of efficient gates in the $i_{th}$ layer of each half circuit, $d$ the depth, $m$ corresponds to the number of the equivalent half-circuits, $t$ is the average time per gate, and $s$ is the number of nodes (or parallel units if more than one node are packed into a unit).

Under the rules described in Ref. [1], we estimate the number of X and Y gates in each layer of each half circuit (expect for the first three layers) to be 6, 8 and 10 for 56-, 64- and 72-qubit circuit, respectively. As we optimize the simulation of diagonal gates, all of the CZ gates and T gates in each half circuit can be combined into 2 gates. Therefore, for all scales of the circuits, we make $n_1=1$, $n_2=2$ and $n_3=2$ while for $i>3$ we let $n_i$ to be 8, 10 and 12 for the 56-, 64- and 72-qubit half circuit, respectively.

## 2.3 Simulation Schemes on Different Hardware Circumstances

When the qubit count (simulated directly) increases from 28 to 32, and further to 36, different simulation strategies should be applied to adapt to the huge difference in the amount of storage required. Here, we propose three possible hardware configurations when simulating a 64-qubit circuit, as well as discuss their schemes and provide formula for estimates of times.

### 2.3.1 Each node has sufficient RAM to simulate a partition.

We perform 56- and 42-qubit circuit simulations under this scheme. In this instance, all copies have to be simulated independently. In total, 65536 copies are simulated. As for each partition, time is required for memory allocation and the simulation of all 22 layers in the half circuit. Also, after sampling the data and performing the tensor product, the results are finally added to the resultant vector. Regarding memory, frequency, and bandwidth, the cores of the CPU are considered the main time consuming factor, and this is the reason why we use a GPU as a computing device for the 56-qubit circuit. The total duration time for a simulation would be $T \times 22 \times 65536/N$, where $T$ represents the time for each layer and $N$ the node number. We note that $N$ should be less than 65536.

*2.3.2 Each node has sufficient RAM to simulate a partition and save the result of the first 14$^{th}$ layers.*

We performed a 64-qubit circuit simulation under this scheme. For the first 14 layers, cutting off 8 CZ gates generates 256 copies. We substituted the simulation for the first 14 layers by preparing them beforehand. The simulation of the last 8 layers then needs the initial state to be loaded locally from RAM. The initial state data is 128 GB per copy.

Assuming the bandwidth for loading the memory with 128 GB of data is $v$ GB/s, the total duration time for simulation is given by $\left(T_\text{p} + T \times 8 + \dfrac{128}{v}\right) \times 256/N$, where $T$ represents the time for each layer, $N$ the node number, and $T_\text{p}$ the time for preparation. Again, note that $N$ should be less than 256.

*2.3.3 Each node does not have sufficient RAM to simulate a partition, and hence the 32-qubit circuit is simulated in the distributive mode.*

We assume $n$ nodes are required to simulate 32 qubits. The upper and lower parts in one circuit can be simulated in the same $n$ nodes. The tensor product of the upper and lower sampled results is also calculated in the distributive mode. First, we copy the sampled result to all $n$ nodes; next, at each node, the components they possess are calculated and then added to the final results. If we pack $n$ nodes into a unit, the total time is calculated identically as for case 1 but by replacing $N$ with the unit count.

*2.4 Methods to optimize simulation of diagonal gates*

The bottleneck of our simulations is not performing calculations, but the speed of reading and writing to memory. It requires $2^n$ reads and writes to memory, and only $2^{n+1}$ floating-point operations for an X-gate of a $n$ qubits system. However, floating-point operations are much faster than memory reads. This was proved in our experiment of 42-qubits quantum random circuit calculated using a GTX-1080Ti. For the GTX-1080Ti, the speed of a single floating-point operation is 32 times faster than the double floating-point operation. For the same random circuit and the numbers of layers, we stored the vectors of the quantum state with single floating-point and double floating-point numbers to compare the time spent in the calculations. The result shows that using single floating-point storage is only twice as fast as double floating-point storage, far less than the above factor of 16. Therefore, to reduce computation times, reducing the number of reads and writes to memory is required as well as optimizing the efficiency of memory reads.

A diagonal gate, such as a CZ, T gate, or a $P_0$, $P_1$ gate as mentioned, can be written in accordance with its definition as a diagonal matrix in the computation basis. Within the simulation process, performing a diagonal gate means only to multiply each component of the state vector with a corresponding diagonal coefficient. This process does not cause data exchange between the two elements in the array.

Inspired by these two factors, we produced a combination of all diagonal gates in one layer. First, the tensor product of two or more diagonal matrices should always be a diagonal matrix. If a traverse through the $2^{32}$ matrix is done when operating an arbitrary diagonal gate, $N$ diagonals combining to form another diagonal gate (even if we cannot calculate the exact form of it) will only make $2^{32}$ queries of memory rather than $N \times 2^{32}$ times.

Specifically, we apply $m$ T gates to the $t_{\text{th}}$ layer of an $n$ qubits quantum system. For convenience in description, we do not consider an instance of circuit partitioning. Let $V_t^{\text{T}} = \{T_1, ..., T_m\}$ be the set of operation qubits of the T gates. Using the properties of diagonal matrices, the final quantum state $\psi_{i_1,...,i_n}^t$ can be written

$$\psi_{i_1...i_n}^t = \sum_{j_1,...j_n \in \{0,1\}} \left( \prod_{k \in V_t^T} T_{i_k,j_k}^{t,k} \psi_{j_1...j_n}^{t-1} \right) = \prod_{k \in V_t^T} T_{i_k,i_k}^{t,k} \psi_{i_1...i_n}^{t-1}. \tag{3}$$

The representations of $T_{i_k,i_k}^{t,k}$ and $CZ_{i_h,i_k,j_h,j_k}^{t,k}$ are given in the Supplementary data.

Considering $T_{0,0}^{t,k} = 1, T_{1,1}^{t,k} = \frac{1+i}{\sqrt{2}}$, Eq. (3) can be rewritten as

$$\psi_{i_1...i_n}^t = \left( \frac{1+i}{\sqrt{2}} \right)^{m_1} \psi_{i_1...i_n}^{t-1}. \tag{4}$$

For each element of this state, $m_1(V_t^{\text{T}})$ is the number of $|1\rangle$ among $V_t^{\text{T}}$.

When using Eq. (4), we only need $2^n$ read and write calls from memory to get the final quantum state. The additional floating-point operations and judgments compared with the time required for memory reads can be neglected. For the time cost, applying $m$ T gates is almost exactly the same as one T gate applied to the same qubit system.

We use the same method in dealing with the CZ gates among $E_{1,t}$ and $E_{2,t}$.

Analogously, for convenience of description, we apply $m$ CZ gates to the $t_{th}$ layer of an $n$ qubits quantum system without considering the case of circuit partition. Let $E_t$ be the set of CZ gates, and let $h(k)$ represent the control (target) qubit. Exploiting the properties of diagonal matrices, the final quantum state $\psi^t_{i_1,\ldots,i_n}$ can be written as

$$\psi^t_{i_1,\ldots,i_n} = \sum_{j_1,\ldots,j_n \in \{0,1\}} \left( \prod_{(h,k)\in E_t} CZ^{t,(h,k)}_{i_h,i_k,j_h,j_k} \psi^{t-1}_{j_1\ldots j_n} \right) = \prod_{(h,k)\in E_t} CZ^{t,(h,k)}_{i_h,i_k,i_h,i_k} \psi^{t-1}_{i_1\ldots i_n}. \quad (5)$$

As $CZ^{t,k}_{0,0,0,0} = CZ^{t,k}_{0,1,0,1} = CZ^{t,k}_{1,0,1,0} = 1, CZ^{t,k}_{1,1,1,1} = -1$, Eq. (5) can be expressed as

$$\psi^t_{i_1,\ldots i_n} = (-1)^{m_2} \psi^{t-1}_{i_1\ldots i_n}. \quad (6)$$

For each element of this state, $m_2(E_t)$ is the number of edges among $E_t$ for which the vertices are both $|1\rangle$.

## 3. Results

In order to characterize the degree of optimization, we measured the relative computing complexity of a task with a number of equivalent qubits ($N_e$) which takes the same RAM to store qubit state vectors as all copies of the original circuit would take. Taking the above eight-qubit circuit as an example. After the splitting of two CZ gates, we obtain eight four-qubit circuits, so $N_e=4+\log_2 8=7$. In short, when a CZ gate is transformed, the number of copies is doubled, which is equivalent to adding an equivalent qubit. To describe circuit compression, we define two more types of qubit number. Real qubit number ($N_r$) is the number of qubits in the original quantum circuit and maximum qubit ($N_m$) is the maximum number of qubits that a computing device can store. We discuss the data compression based on a simulation of a 56-qubit circuit calculated by a single PC with a 256 GB RAM. Here, $N_r=56$, $N_m=33$. The first six layers divide the circuit into two sub–circuits, each of which has 28 qubits without splitting any CZ gate, hence $N_e=29$. After that, the splitting of each CZ gate results in adding one equivalent qubit. When $l\leq 7$ ($l$ represents the number of layers), $N_e<N_m<N_r$, which means our method can get vectors of all quantum states by saving the result of each sub-circuit. As $8<l\leq 38$ and hence $N_m<N_e<N_r$, we can still compress circuit information effectively, but we have to sample the result and cannot get the full vectors of quantum states. This is equivalent to a lossy compression. As $l>38$ and hence $N_m<N_r<N_e$, our method fails to compress the circuit, since the time and space complexity are used in exchange of the feasibility of simulation.

As an example of a 56-qubit simulation with different circuit depth, Table 2 presents the relationship between the number of transformed CZ gates and its relative complexity. We find the relative complexity of this algorithm is proportional to $O(\frac{n}{2}+c)$, with $n$ being the total number of qubits. In Fig. 2a, we plot the relationship between the number of qubits (topologies), the depth of the circuit, and the relative complexity. The complexities increase only when the 7$^{th}$ and 8$^{th}$ layers occur in one loop. The dashed lines indicate that the optimization has an upper bound for a given depth, beyond which the partition scheme does not help to reduce the complexity. In Fig. 2b, we plot the relationship between runtime and depth for 6×6, 6×7, 7×7 and 8×7, random circuits simulated by a PC with GTX-1080Ti, respectively. When the depth is large, we use the time cost of several copies to calculate the total runtime. The results show that the runtime is positively correlated with the complexity, as they both increase with circuit depth.

Determining how the qubits are split is a combinatorial optimization problem. We do not completely solve this problem. However, three different partitioning schemes are demonstrated in Fig. 3. Roughly, the relative complexity is proportional to $O(\frac{n}{t}+c_t)$, where $t$ is the number of dividing parts and $c_t$ is number of corresponding transformed CZ gates. $c_t$ depends on the specific partitioning scheme, but generally increases linearly as $t$ increases. As the data show, dividing the circuit into three or four parts is more effective if the circuit depth is small. However, dividing the circuit into two parts is more suitable for depths around 22 or 30.

Using the scheme to simulate a 72-qubit circuit with a depth less than 30 is also possible in our estimation. Here, we suppose that this task is done with an identical supercomputer on which the IBM team simulated a 56-qubit circuit. The supercomputer has 24576 nodes, each equipped with 16 cores and 16-GB of RAM. We calculate an average gate time of 0.67 s for a 36-qubit circuit with a 128-nodes cluster, in accordance with the previous work [18]. In these circumstances, we estimate that 16 h are needed to perform the simulation of a 72-qubit circuit of depth 23 within 228 sampling number.

Because the number of simulated qubits directly decreases, we designed experiments to simulate 8×8-, 8×7-, and 6×7-qubit topologies with far fewer hardware resources than those used by the IBM team to obtain their numerical results. All circuits have depth 22 (see Fig. 4 for illustrations). These circuits are constructed under the rules described in Ref. [6]. We calculate the probability amplitude of the sampled quantum-state components and plot the distribution of the probabilities using MATLAB (Fig. 5). The figure shows the log-transformed outcome probabilities, plotted together with their theoretical distribution, which is given by the Gumbel distribution:

$$f(z)=\frac{1}{\alpha}\exp\left(z-\frac{1}{\alpha}\left(e^z+\alpha-1\right)\right), \quad (7)$$

with $\alpha=1$. The distributions obtained in the 56- and 64-qubit simulations are all matched well to the theory. Data sampled from the full vectors of the final states exhibit a Porter–Thomas distribution, as theory predicts. Under our scheme, an arbitrary number of measurement outcomes of the full vector space can be extracted without extra cost. We sampled $2^{28}$ elements in 64- and 56-qubit simulations. In the 42-qubit simulation, we obtained distributions with $2^{20}$ and $2^{28}$ sampling data, suggesting that sampling number does not affect accuracy. Table 3 presents the experimental environments and time costs (see Supplementary data for more details). The time costs only include the calculation times. The times for data transfer and for figure plotting is not considered.

## 4. Conclusion

The IBM team simulated a 56-qubit circuit with a depth of 23 using the supercomputer mentioned above. Because the 28-qubit state vector occupies 4-GB of RAM, the 28-qubit circuit can be simulated within one node; each gate would then take 0.25 s on average to simulate [18]. In these circumstances, we estimate that the same circuit can be simulated in 460 s, a much shorter time than that achieved by the IBM team. Similarly, if circuits of depth 30, 31, and 38 were simulated, we estimate the required times to be 2.74, 22.7 h, and 18.9 d, respectively.

The Google team also studied the Feynman path method to be applied in the simulation [8]. An obvious characteristic of this method is that each run of the calculation obtains one measurement outcome, and their team simulated 200 thousand components of a 56-qubit circuit. The cost of their algorithm is exponential of order $\min(O(dl), O(n))$ for depth $d$, minimum lateral dimension $l$, and total number of qubits $n$. However, our scheme is not affected by the sample number, which is suitable for acquiring a great number of outcomes. In our experiments, we obtained $2^{28}$ components for the 56- and 64-qubit circuit simulations. According to our experimental results, our scheme is optimal when extracting more than one million outcomes.

Simulating more than 49 qubits has no conflict with the quantum supremacy. The optimization methods reduce the memory but introduce new variables (e.g., the number of CZ gates for partitioning method, the number of non-diagonal gates for the Feynman path method). Due to the complexity growing exponentially with those variables, the simulation of more than 50 qubits will always have an upper bound in depth. Nevertheless, the simulation of more qubits system with small depth still plays an important role for the study of quantum algorithms such as QFT and unsupervised machine learning [19]. Moreover, the partitioning scheme could be combined with other simulation methods (e.g., Feynman path integral), to further reduce the complexity. These improvements may help to realize the simulation of many other quantum algorithms.

## Conflict of interest

The authors declare that they have no conflict of interest.

## Acknowledgments

This work was supported by the National Key Research and Development Program of China (2016YFA0301700), the National Natural Science Foundation of China (11625419), and the Anhui Initiative in Quantum Information Technologies (AHY080000). The numerical computations were supported by Yangzi Cloud Computing Data Centre and Gyrotech, Nanjing China.

## Author contributions

All authors contributed extensively to the work presented in the paper; Z.-Y.C. and Q.Z. wrote the simulation programs; Z.-Y.C., X.Y., G.-C.G. and G.-P.G. conducted theoretical analysis; Z.-Y.C., Q.Z. and C.X. wrote the data analysis programs and performed data analysis; Z.-Y.C., Q.Z. wrote the manuscript and G.-P.G. supervised the project.

## Data availability

The data that support the plots within this paper and other findings of this study are available from the corresponding author upon reasonable request.

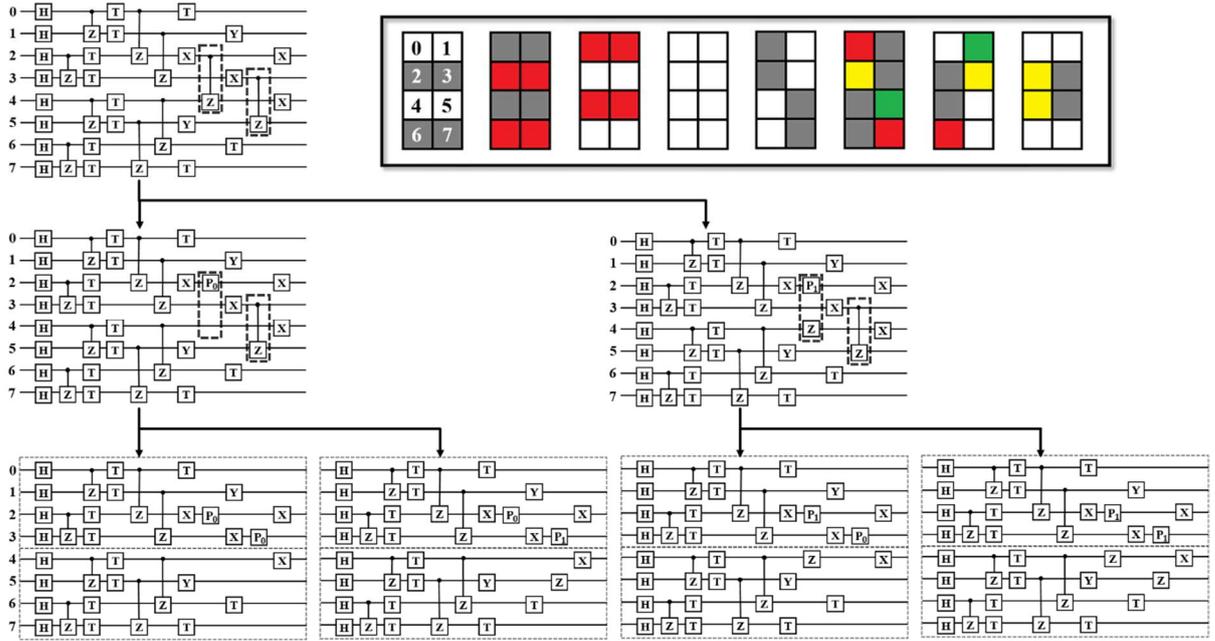

Fig. 1 An example for partitioning. The first row is the original circuit. The two CZ gates in the dashed boxes entangle the first and last four qubits. Next, the left CZ gate is transformed, the original circuit being equivalent to the addition of the circuits in the second row. Continuing, the right CZ gate is transformed, generating four circuits in the third row. The final state of the original circuit is equal to the addition of all transformed circuits. The dashed boxes in the third line divide each circuit into two parts, where they can be simulated independently. Inset: The 4×2 quantum circuit of depth 8 expressed in the form of a grid, where the coloured squares represent different gates: yellow–an X gate, green–a Y gate, red–a T gate, and grey–a CZ gate. Each graph represents a layer. The numbers denote the corresponding qubit in this example.

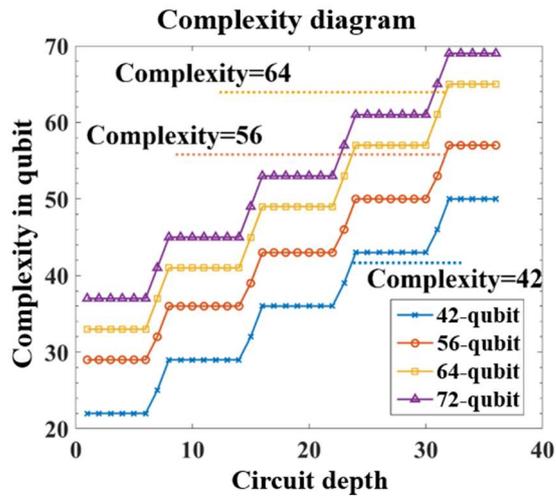 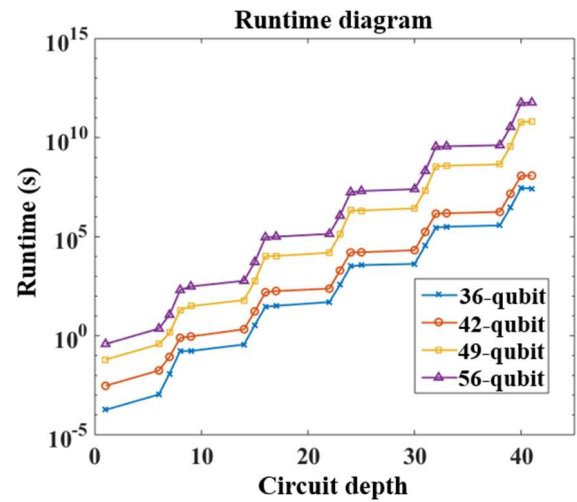

Fig. 2 Complexity and runtime diagram with changes in depth and qubit counts. (a) The violet triangle, yellow square, orange circle and blue cross lines correspond to the simulation complexity of the 72-, 64-, 56-, and 42-qubit universal random circuits, respectively. The dotted lines link the computation complexity of the original circuits where the upper bounds of the optimization are. (b) The four solid lines correspond to runtime of 6×6, 6×7, 7×7 and 8×7 universal random circuits simulated by a PC with GTX-1080Ti, respectively. Runtime is positively correlated with complexity.

(a)

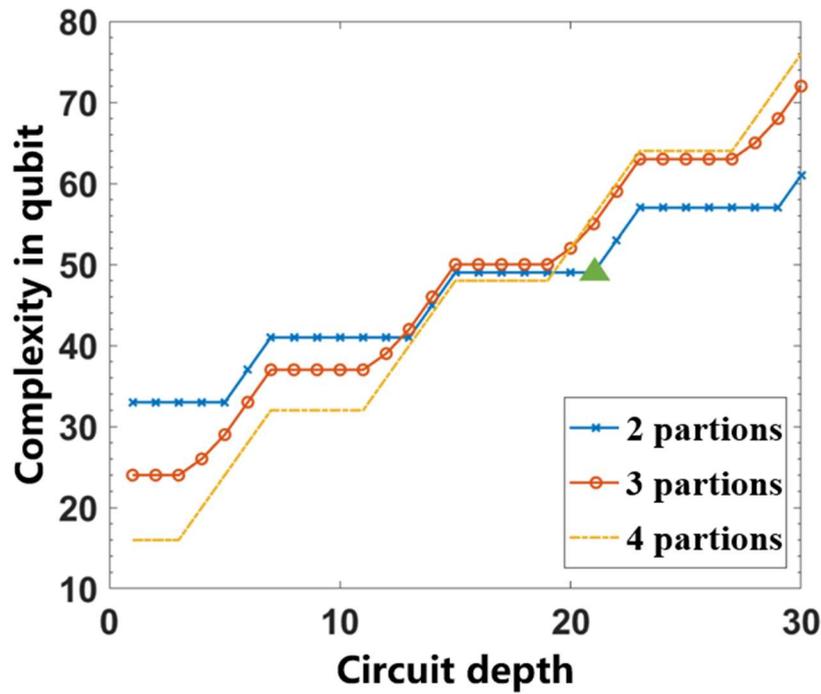

(b) (c) (d)

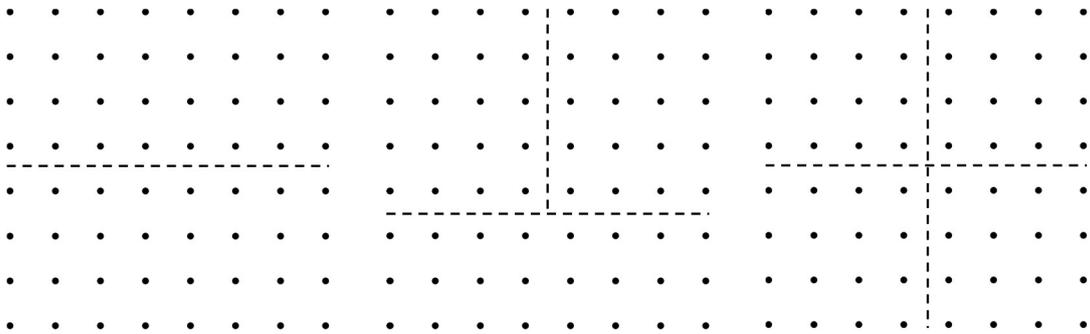

Fig. 3 Three schemes for circuit partitioning, and the relative complexity as a function of circuit depth. (a) The line chart for complexity versus circuit depth for the three schemes shown in (b), (c), and (d). The "complexity in qubit" means the relative calculated complexity for both time and space is equivalent to that qubit count. We calculated 22 layers, dividing the circuit into two parts (marked by the green triangle), which is equivalent to a 49-qubit circuit simulation using the direct approach. (b) The 64-qubit circuit is divided into two parts and corresponds to the blue line with dotted crosses in (a); (c) The 64-qubit circuit divided into three parts and corresponds to the red line with circles in (a). The partition with the maximum number of qubits should be counted as the relative simulation complexity, which is the lower part having 24 qubits;

(d) The 64-qubit circuit divided into four parts and corresponds to the yellow dot line in (a).

(a)

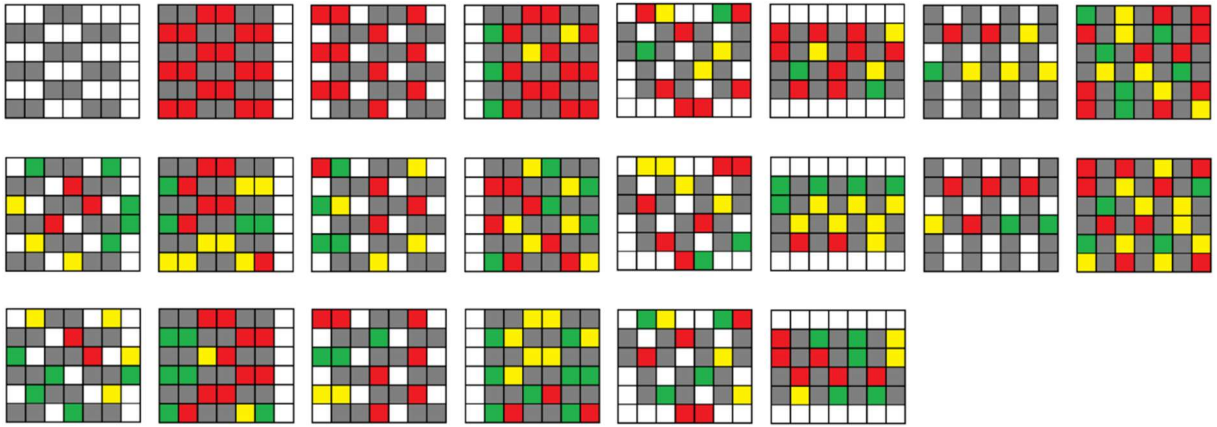

(b)

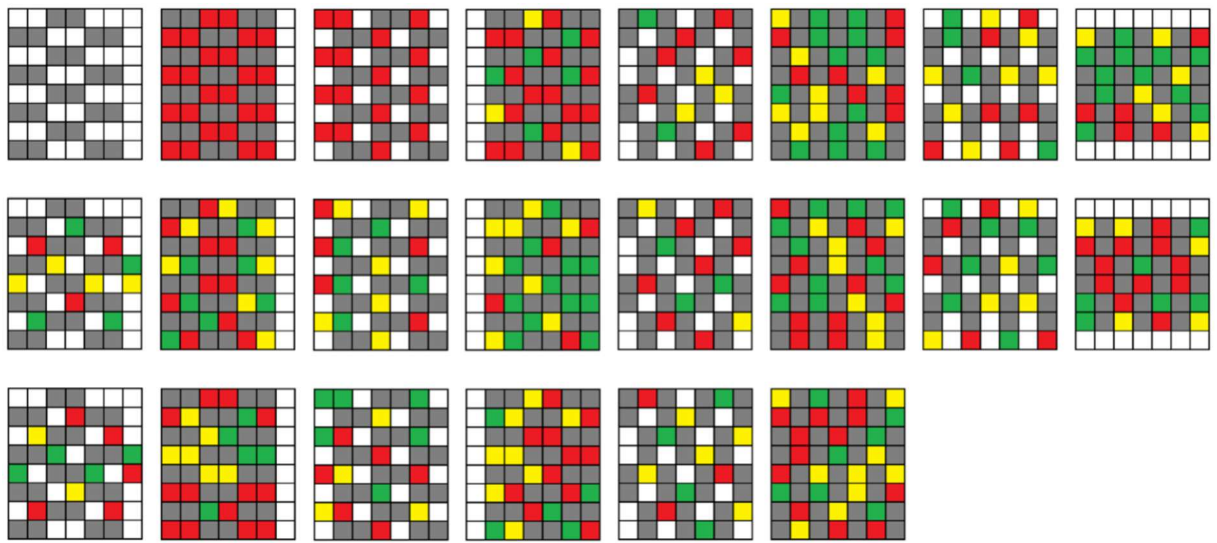

(c)

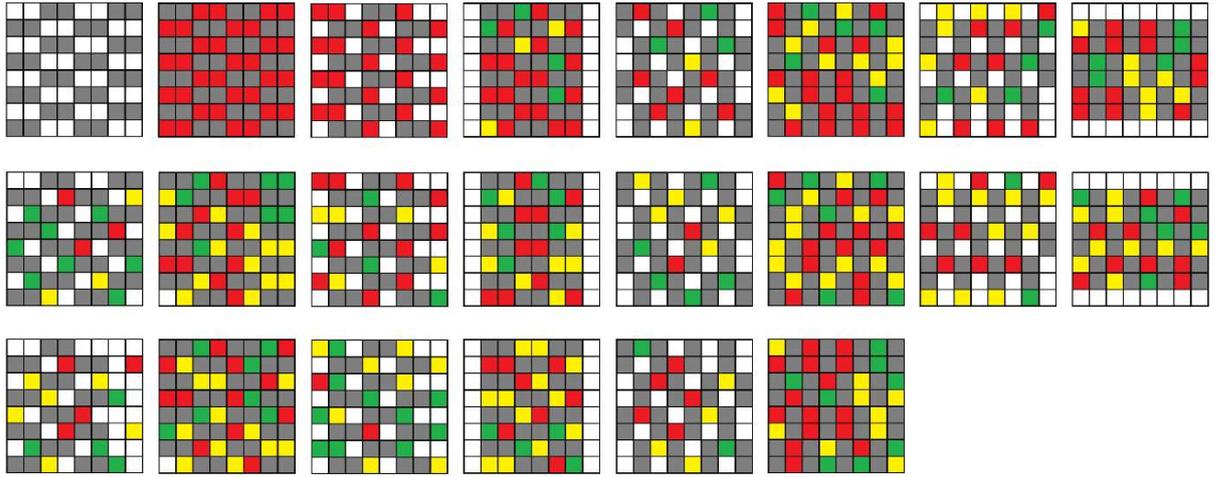

Fig. 4 42-, 56-, and 64-qubit universal random circuit. The representation of the circuits follows that in Fig. 1. (a) 42-qubit circuit with depth 22. There are in total 494 gates comprising 80 X gates, 75 Y gates, 147 T gates, and 192 CZ gates; (b) 56-qubit circuit with depth 22. There are in total 682 gates comprising 105 X gates, 119 Y gates, 188 T gates, and 270 CZ gates; (c) 64-qubit circuit with depth 22. There are in total 784 gates comprising 146 X gates, 99 Y gates, 225 T gates, and 312 CZ gates.

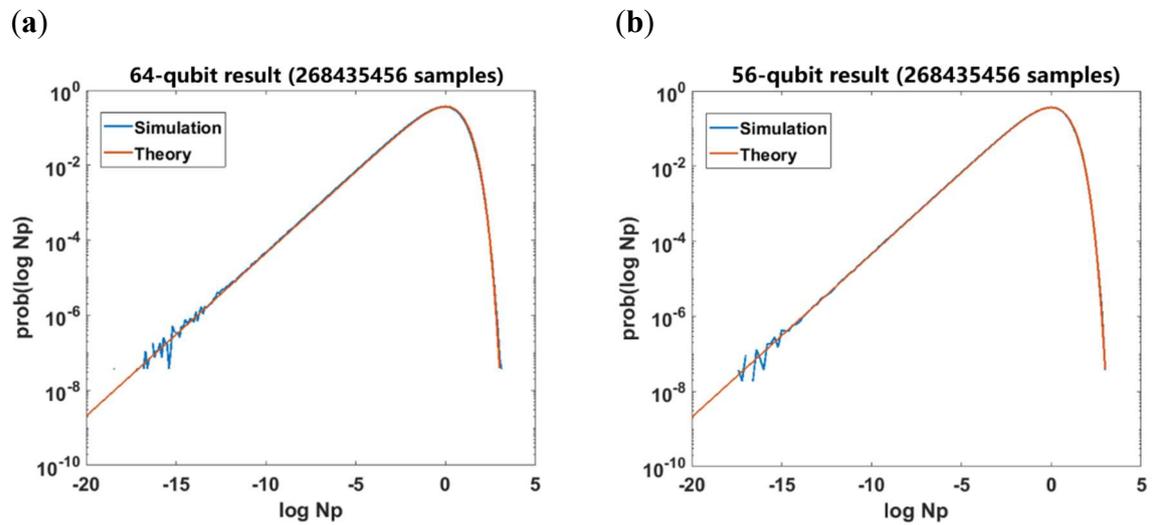

(c)

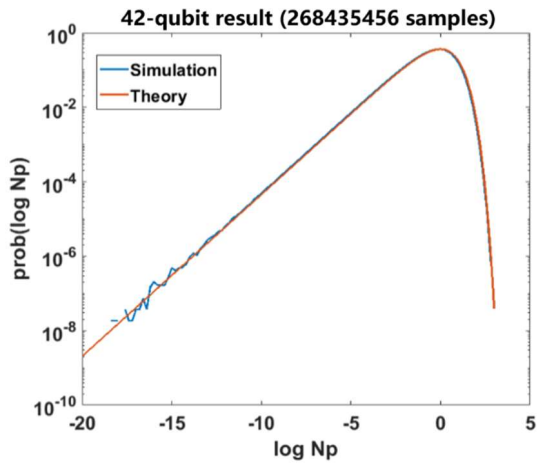 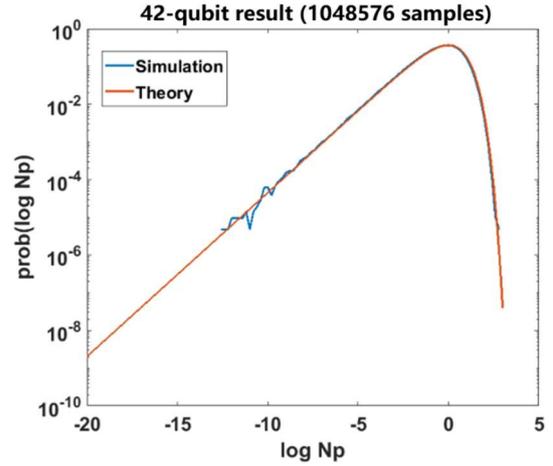

(d)

Fig. 5 Plots of the log-transformed measurement outcome probabilities of 42-, 56- and 64-qubit simulation, plotted with the theoretical distribution. Simulation results of (a) the 64-qubit circuit, (b) the 56-qubit circuit, and (c) the 42-qubit circuit, sampling 268435456 outcomes from the final results; (d) Same as in (c), but sampling 1048576 outcomes. Comparing (c) with (d), the sampling count does not affect the simulation distribution.

Table 1 Time estimations for different qubit counts and depths for the universal random circuit. The average time per gate of the 28-qubit circuit is estimated to be 0.25 s, and that of the 32-qubit and 36-qubit circuits (obtained from Ref. [18]) are 0.38 and 0.67 s, respectively.

| Qubit | Depth | Time estimation |
|---|---|---|
| 56 | 22 | 52.3 s |
|  | 23 | 7.33 min |
|  | 30 | 2.62 h |
|  | 31 | 21.7 h |
|  | 38 | 18.0 d |
|  | 39 | 148 d |
| 64 | 22 | 6.59 min |
|  | 23 | 1.85 h |
|  | 30 | 1.65 d |
|  | 31 | 27.4 d |
| 72 | 22 | 55.5 min |
|  | 23 | 15.6 h |

Table 2 Number of CZ gates to be transformed and the relative simulation complexity in qubits when simulating a 56-qubit circuit.*: We simulated a circuit of depth 22.

| Depth | Number of CZ gates to be transformed | Total number of CZ gates to be transformed | Total number of equivalent 28-qubit circuits | Relative simulation complexity in qubits |
|---|---|---|---|---|
| 1–6 | 0 | 0 | $2^1=2$ | 29 |
| 7 | 3 | 3 | $2^4=16$ | 32 |
| 8–14 | 4 | 7 | $2^8=256$ | 36 |
| 15 | 3 | 10 | $2^{11}=2048$ | 39 |
| 16–22* | 4 | 14 | $2^{15}=32678$ | 43 |
| 23 | 3 | 17 | $2^{18}=262144$ | 46 |
| 24–30 | 4 | 21 | $2^{22}=4194304$ | 50 |
| 31 | 3 | 24 | $2^{25}=33554432$ | 53 |
| 32–38 | 4 | 28 | $2^{29}=536870912$ | 57 |

Table 3 Details of the qubit circuit simulations

| Qubit | Topology | Depth | Hardware | Software | Time of simulation | Average time per gate |
|---|---|---|---|---|---|---|
| 42 | 6×7 | 22 | GTX-1080Ti * 1 | CUDA | 0.384 h | 0.00017 s |
| 56 | 8×7 | 22 | GTX-1080Ti * 1 | CUDA + OpenMP | 44.1 h | 0.014 s |
| 64 | 8×8 | 22 | Xeon E5-2690V4 (14 cores) * 128 | OpenMP | 31.3 h | 0.28 s |

# Supplementary Information

## A. Theoretical derivation of qubit partitioning

As the 64-qubit system needs a lot of space to store the vector of the quantum states, we divided the circuit into two independent parts so that we can calculate them sequentially. From Figure 1a, the upper part of the circuit is clearly connected with the lower part of the circuit by the CZ gate at the last two layers for a circle of eight layers. Therefore, if we can divide CZ gate into several single gates, the scale of the calculation is greatly reduced.

First, an arbitrary two-qubit control gate is decomposed as a sum of two tensor products of $2 \times 2$ matrices. Let $U^{ij}$ be the two-qubit control gate applied to qubits $i$ and $j$, where qubit $i$ is the controller, and qubit $j$ is the target qubit; $U^{ij} \in C^{4 \times 4}$. A two-qubit control gate means when qubit $i$ is in the state $|0\rangle$, qubit $j$ does nothing, and qubit $i$ is in the state $|1\rangle$, a single qubit gate is performed to qubit $j$, denoted as $U^j$. Hence, $U^{ij}$ can be re-expressed as:

$$U^{ij} = P_0^i \times I^j + P_1^i \times U^j \tag{1}$$

where

$$P_0 = \begin{pmatrix} 1 & 0 \\ 0 & 0 \end{pmatrix}, P_1 = \begin{pmatrix} 0 & 0 \\ 0 & 1 \end{pmatrix},$$

and $I$ represents the unit matrix. For a CZ gate, it can be re-expressed as:

$$CZ^{ij} = P_0^i \times I^j + P_1^i \times Z^j \tag{2}$$

where $Z$ is a Pauli-Z gate expressed as

For convenience in description, we describe the quantum random circuit the same way as [7]. We define a circuit to be a sequence of $m$ layers $G_1,...,G_m$, where each layer consists of an undirected graph on the same vertex set $V := 1,...,n$; formally, $G_t = (V_t, E_t)$, $V_t = 1,...,n$, $E_t \subseteq V_t \times V_t$. Every edge represents a CZ gate, and no vertex can appear in two edges on the same layer. At each layer, a single-qubit gate is applied to every vertex that does not appear in a two-qubit gate.

Let

$$V_t^1 = V_t \left\{ j : (i,j) \in E_t \text{ or } (j,k) \in E_t \right\} \tag{3}$$

be the qubit in layer $t$ that is not involved in the two-qubit gates. For $j \in V_t^1$, let $U^{t,j}$ be the gate applied to qubit $j$ in the layer $t$. For $(i,j) \in E_t$, let $U^{t,(i,j)}$ be the two-qubit gate applied to qubit $i$ and $j$ in layer $t$.

We use subscripts to denote elements of vectors and matrices, and in particular we use binary strings as indices as is customary in quantum mechanics.

Let $n$ be the number of qubits, and for our circuit, we let $n = r \times q$ where $r$ represents the row number of the vertex, and $q$ represents the column number of the vertices for each layer; At least one of the $r$'s and $q$'s is an even number in order to divide the circuit equally. However, even if both $r$ and $q$ are odd, the method we state below may also be applied. For convenience in expression, we let $r$ be an even number $r = 2p$.

Using the above notation, and given the state $\psi_{t-1}$ after the $(t_1)$th layer of gates, we can express the state after the $t$th layer as:

$$\psi_{i_1 \ldots i_n}^t = \sum_{j_1, \ldots, j_n \in \{0,1\}} \left( \prod_{k \in V_t^1} U_{i_k, j_k}^{t,k} \prod_{(h,k) \in E_t} U_{i_h i_k, j_h j_k}^{t,(h,k)} \psi_{j_1 \ldots j_n}^{t-1} \right) \tag{4}$$

Now, by dividing the circuit into two half-circuits along the center line, we get $V_t = \{V_{1,t}, V_{2,t}\}$, where $V_{1,t} = \{1, \ldots, p*q\}$ and $V_{2,t} = \{p*q+1, \ldots, 2p*q\}$.

$E_t = \{E_{1,t}, E_{2,t}, E_{int,t}\}$, where $E_{1,t} \subseteq V_{1,t} \times V_{1,t}$, $E_{2,t} \subseteq V_{2,t} \times V_{2,t}$, and $E_{int,t} \subseteq V_{1,t}^2 \times V_{2,t}^2$,

where $V_{1,t}^2 = \{(p-1)*q+1, \ldots, p*q\}, V_{2,t}^2 = \{p*q+1, \ldots, (p+1)*q\}$.

Also, we let

$$V_{1,t}^1 = V_t \left\{ j : (i,j) \in E_{1,t} \text{ or } (j,k) \in E_{1,t} \right\}, V_{2,t}^1 = V_t \left\{ j : (i,j) \in E_{2,t} \text{ or } (j,k) \in E_{2,t} \right\}.$$

According to the pattern of the CZ-gate location in Figure 5, we find that only in layer $a \times q + 7$ and $a \times q + 8$ (where $a$ is an integer) is $E_{int,t}$ non-empty. For the first six layers ($t < 7$), we rewrite (4) meaning that the first six layers can be divided into two half-circuits directly.

$$\psi_{i_1 \ldots i_n}^t = \sum_{j_1, \ldots, j_{p \times q} \in \{0,1\}} \left( \prod_{k \in V_{1,t}^1} U_{i_k, j_k}^{t,k} \prod_{(h,k) \in E_{1,t}} U_{i_h i_k, j_h j_k}^{t,(h,k)} \psi_{j_1 \ldots j_{p \times q}}^{t-1} \right)$$

$$\sum_{j_{p \times q+1}, \ldots, j_{2 p \times q} \in \{0,1\}} \left( \prod_{k \in V_{2,t}^1} U_{i_k, j_k}^{t,k} \prod_{(h,k) \in E_{2,t}} U_{i_h i_k, j_h j_k}^{t,(h,k)} \psi_{j_{p \times q+1} \ldots j_{2 p \times q}}^{t-1} \right) \tag{5}$$

For the 7th layer ($t=7$),

$$\psi^t_{i_1,\ldots i_n} = \sum_{j_1,\ldots j_n \in \{0,1\}} \left( \prod_{k \in V^1_{1,t}} U^{t,k}_{i_k,j_k} \prod_{k \in V^1_{2,t}} U^{t,k}_{i_k,j_k} \prod_{(h,k) \in E_{1,t}} U^{t,(h,k)}_{i_h i_k, j_h j_k} \prod_{(h,k) \in E_{2,t}} U^{t,(h,k)}_{i_h i_k, j_h j_k} \prod_{(h,k) \in E_{int,t}} U^{t,(h,k)}_{i_h i_k, j_h j_k} \psi^{t-1}_{j_1 \ldots j_n} \right)$$
(6)

Here, to help in describing the formula, we let $P^{t,j}_0 (P^{t,j}_1)$ be the $P$ gate applied to qubit $c$ in layer $t$, $Q^{t,j}_0 (Q^{t,j}_1)$ be the $I(Z)$ gate applied to qubit $j$ in layer $t$. Using (2), we can divide the CZ gate in $E_{int,t}$ and obtain (7)

$$\psi^t_{i_1,\ldots i_n} = \sum_{j_1,\ldots j_n \in \{0,1\}} \prod_{k \in V^1_{1,t}} U^{t,k}_{i_k,j_k} \prod_{k \in V^1_{2,t}} U^{t,k}_{i_k,j_k} \prod_{(h,k) \in E_{1,t}} U^{t,(h,k)}_{i_h i_k, j_h j_k} \prod_{(h,k) \in E_{2,t}} U^{t,(h,k)}_{i_h i_k, j_h j_k}$$
$$\left( \sum_{l_m \in \{0,1\}, m \in V^2_{1,t}} \prod_{k \in V^2_{1,t}} P^{t,k}_{l_m,i_k,j_k} \prod_{k \in V^2_{2,t}} Q^{t,k}_{l_m,i_k,j_k} \right) \psi^{t-1}_{j_1 \ldots j_n}$$
$$= \sum_{l_m \in \{0,1\}, m \in V^2_{1,t}} \sum_{j_1,\ldots j_n \in \{0,1\}} \left( \prod_{k \in V^1_{1,t}} U^{t,k}_{i_k,j_k} \prod_{(h,k) \in E_{1,t}} U^{t,(h,k)}_{i_h i_k, j_h j_k} \prod_{k \in V^2_{1,t}} P^{t,k}_{l_m,i_k,j_k} \right)$$
$$\left( \prod_{k \in V^1_{2,t}} U^{t,k}_{i_k,j_k} \prod_{(h,k) \in E_{2,t}} U^{t,(h,k)}_{i_h i_k, j_h j_k} \prod_{k \in V^2_{2,t}} Q^{t,k}_{l_m,i_k,j_k} \right) \psi^{t-1}_{j_1 \ldots j_n}$$
(7)

Now, there is no two-qubit gate in $E_{int,t}$. In addition, the final quantum state of the $6_{th}$ layer splits into two half-quantum states in the form of a tensor product, which means in the $7_{th}$ layer, the upper-circuit and the lower-circuit can be calculated independently. Also the final quantum state $\psi^t_{j_1 \ldots j_n}$ may be written as (8)

$$\psi^t_{i_1,\ldots i_n} = \sum_{l_m \in \{0,1\}, m \in V^2_{1,t}} \left( \sum_{j_1,\ldots j_{p \times q} \in \{0,1\}} \left( \prod_{k \in V^1_{1,t}} U^{t,k}_{i_k,j_k} \prod_{(h,k) \in E_{1,t}} U^{t,(h,k)}_{i_h i_k, j_h j_k} \psi^{t-1}_{j_1 \ldots j_{p \times q}} \right) \right.$$
$$\left. \sum_{j_{p \times q+1} \ldots j_{2 p \times q} \in \{0,1\}} \left( \prod_{k \in V^1_{2,t}} U^{t,k}_{i_k,j_k} \prod_{(h,k) \in E_{2,t}} U^{t,(h,k)}_{i_h i_k, j_h j_k} \psi^{t-1}_{j_{p \times q+1} \ldots j_{2 p \times q}} \right) \right)$$
(8)

For the $8_{th}$ layer and remaining layers, we can use the same method to turn the CZ gate in $E_{int}$ into a combination of single-qubit gates, arriving at two completely independent half-circuits of all layers.

However, when the number of CZ gates in $E_{int}$ increases, this method become inefficient. For when we turn a CZ gate into a group of single-bit gate, we add two sub-circuits for the exchange. Therefore, the amount of computation exponentially increases with the number of CZ gates that are segmented. This is why we simulate 22 layers of the circuit. There is no CZ gate in $E_{int,t}$ in layer 16 to layer 22. When the number of layers increases to 23 layers, the amount of computation increases sixteen-fold, which is hard to simulate with our computing resources.

## B. Hardware and software environments

We applied various environment to various topology. It is possible to simulate a quantum random circuit with a personal computer equipped with a graphics card if there are less than 50 qubits. As tests showed, simulation of 42-qubit consumed 1384 seconds and 56-qubit 44.1 hours.

To face the difficulty that 128-GB memory is too much for graphics card, we turned to high performance computer to finish the 64-qubit simulation. We finally use 128 nodes and each has two Xeon E5-2690V4 (14 cores, 35 MB cache, 2.60 GHz) with 256-GB RAM.

We performed another 64-qubit simulation in the different hardware. We use 64-node and each has two Xeon E5-2667V4 (8 cores, 25MB cache, 3.20 GHz) with 128-GB RAM. The simulation took 10 days. In detail, the state after the sixth layer is stored in hard disk, instead of RAM. Hence, the performance is limited by the read speed of the hard disk.

An effective method for programming a small scale random circuits is to employ CUDA as the software environment. If the graphics memory can load the full vector of a multi-qubit system, it is always faster to use GPU than CPU thanks to the high bandwidth of the graphics memory and a great amount of stream processors. 29-qubit system occupied 8GB memory, so that a little bit more than 8GB graphics memory (e.g. GTX-1080Ti with 11GB graphics memory).

C/C++ is used as our programming language. We use Compute Unified Device Architecture (CUDA) when performing the GPU phase. OpenMP is used at 56-qubit simulation, to accelerate sampling process, which is handled by CPU. MPI is used to manipulate many nodes when simulating 64-qubit. However, no inter-process communication is applied in this process. Each partition of our simulation is computed independently among all nodes. No other tools or libraries are applied.

In the program of 56-qubit or less, it is possible to randomly generate quantum circuits with QRunes Code, gates layout shown in qubit topology. Because all diagonal gates in one layer is calculated as a whole, we only need to define where X or Y gates lies in each layer, thus the executed code is generated by a parser program. In the final program of 64-qubit, we use macro definition to define each operation of the quantum circuits, and another program to parse the circuit and generate the code.